\documentclass[12pt]{iopart}

\usepackage{graphicx}
\usepackage{placeins} 
\usepackage{shuffle}

\usepackage[utf8]{inputenc}
\usepackage{amssymb}
\expandafter\let\csname equation*\endcsname\relax
\expandafter\let\csname endequation*\endcsname\relax
\usepackage{bbold}
\usepackage{amsmath}
\usepackage{slashed}
\usepackage{nicefrac}
\usepackage{braket}
\usepackage{hyperref}
\usepackage[nosort]{cite}
\usepackage{etoolbox}

\makeatletter
\def\@mkboth#1#2{}
\newlength\appendixwidth
\preto\appendix{\addtocontents{toc}{\protect\patchl@section}}
\newcommand{\patchl@section}{%
  \settowidth{\appendixwidth}{\textbf{Appendix }}%
  \addtolength{\appendixwidth}{1.5em}%
  \patchcmd{\l@section}{1.5em}{\appendixwidth}{}{\ddt}%
}
\makeatother

\usepackage{tikz}

\usetikzlibrary{calc,arrows, decorations.markings, shapes.misc, decorations.pathmorphing}

\tikzset{
  ->-/.style={
    decoration={
      markings,
      mark=at position #1 with {\arrow{latex}}},
    postaction={decorate}
  },
  ->-/.default=0.5
}

\tikzset{
    wavy/.style={decorate, decoration={snake}, draw=red},
}

\tikzset{VO/.style={cross out, draw, 
         minimum size=5pt, 
         inner sep=0pt, outer sep=0pt}}

\tikzset{VOline/.style={decorate, decoration={snake}},
}

\tikzset{
    partial ellipse/.style args={#1:#2:#3}{
        insert path={+ (#1:#3) arc (#1:#2:#3)}
    }
}

\newcommand{\cN}{\mathcal{N}}

\newcommand{\black}{\color{black}}

\begin{document}

\begin{figure}[ht]
%%\fl
\begin{center}
\scalebox{0.37}{\includegraphics{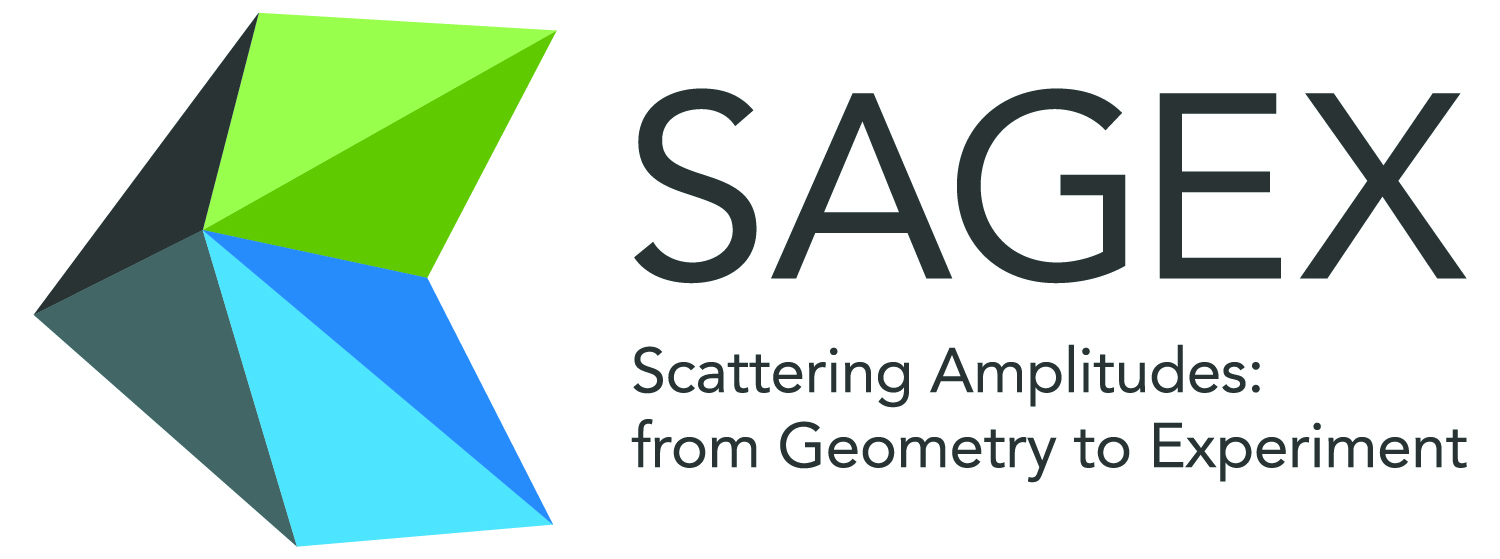}}\qquad \qquad \qquad
%\scalebox{0.035}{\includegraphics{EUflag.jpg}}
\end{center}
\end{figure}
\vspace{-2.7cm}
\begin{flushright}
	SAGEX-22-01\\
	%HU-EP-**/**\\
	%QMUL-PH-22-01
\end{flushright}
\title[The SAGEX Review on Scattering Amplitudes]{The SAGEX Review on Scattering Amplitudes}

\author{Gabriele~Travaglini$^{1}$,   
Andreas~Brandhuber$^{1}$, 
Patrick~Dorey$^{2}$,
Tristan~McLoughlin$^{3, 4}$,
Samuel~Abreu$^{5,6}$,
Zvi~Bern$^{7}$,
N.~Emil~J.~Bjerrum-Bohr$^{8}$,
Johannes~Bl\"{u}mlein$^{9}$,
Ruth~Britto$^{3}$,
John~Joseph~M.~Carrasco$^{10,11}$,
Dmitry~Chicherin$^{12}$, 
Marco~Chiodaroli$^{13}$,
Poul~H.~Damgaard$^{8}$,
Vittorio~Del~Duca$^{14,15,16}$, 
Lance J.~Dixon$^{17}$, 
Daniele~Dorigoni$^{2}$,
Claude~Duhr$^{18}$,
Yvonne~Geyer$^{19}$,
Michael~B.~Green$^{20}$,
Enrico~Herrmann$^{7}$, 
Paul~Heslop$^{2}$,
Henrik~Johansson$^{13,21}$,
Gregory~P.~Korchemsky$^{11,22}$,
David~A.~Kosower$^{11}$,
Lionel~Mason$^{23}$,
Ricardo~Monteiro$^{1}$,
Donal~O’\,Connell$^{6}$,
Georgios~Papathanasiou$^{24}$,
Ludovic~Plant\'{e}$^{8}$,
Jan~Plefka$^{4}$,
Andrea~Puhm$^{25}$,
Ana-Maria~Raclariu$^{26}$,
Radu~Roiban$^{27}$,
Carsten~Schneider$^{28}$,
Jaroslav~Trnka$^{29}$,
Pierre~Vanhove$^{11}$,
Congkao~Wen$^{1}$,
Chris~D.~White$^{1}$ 
}

\begin{figure}[ht]
	\begin{center}
	\begin{eqnarray}
	\includegraphics[scale=0.4]{./Figures/polytope.mps}~~~~~~~~~~
	\includegraphics[scale=0.4]{./Figures/amp.mps}~~~~~~~~~~
	\includegraphics[scale=0.4]{./Figures/exp.mps}
	\nonumber
	\end{eqnarray}
	\end{center}
\end{figure}

\begin{abstract}
This  is an introduction to, and invitation to read, a series of review articles on scattering amplitudes in gauge theory, gravity, and superstring theory. Our aim is  to provide an overview of the field, from basic aspects to a selection of  current (2022)  research and developments.

\end{abstract}

\noindent\begin{minipage}{0.10\textwidth}% adapt widths of minipages to your needs
\includegraphics[width=\linewidth]{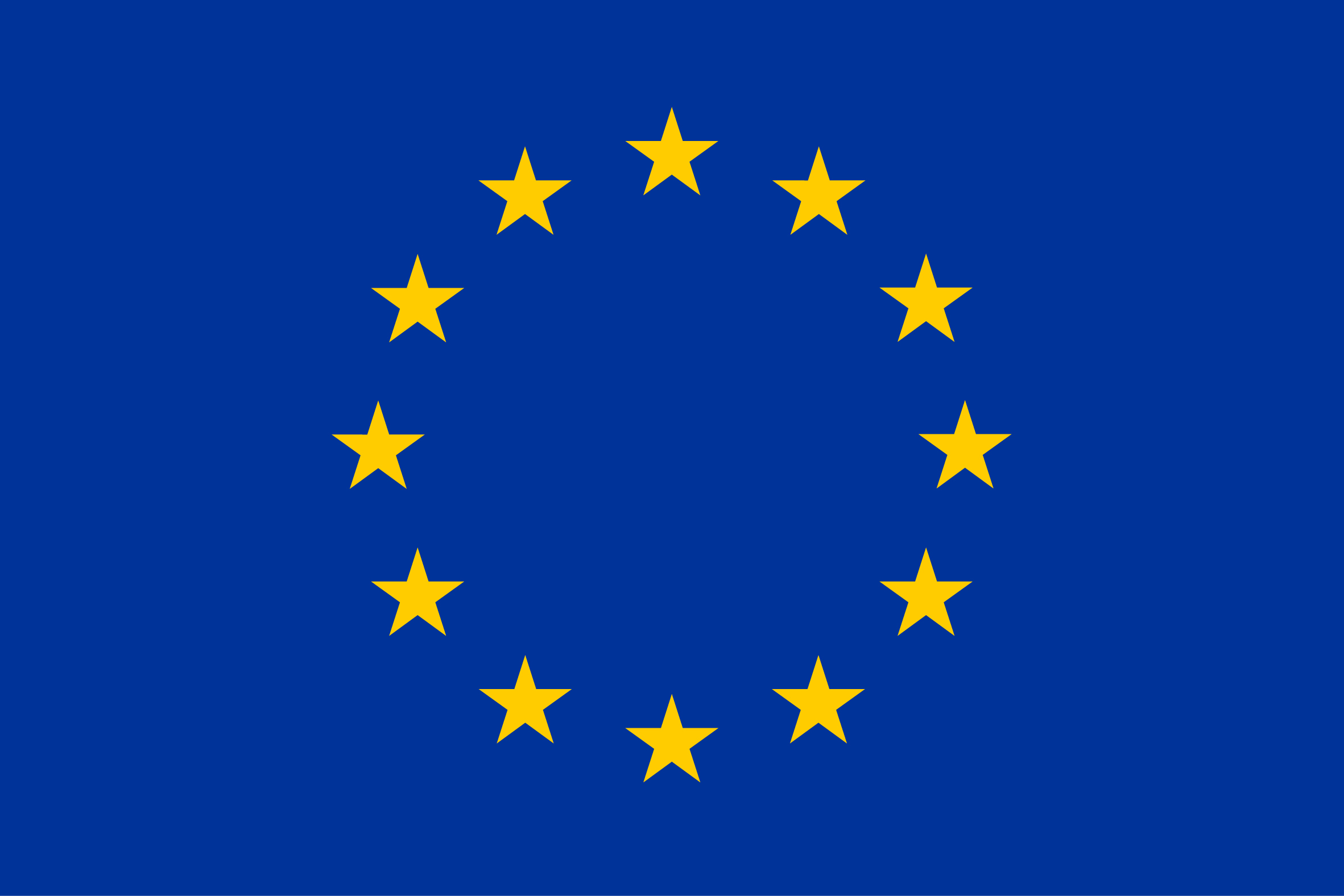}
\end{minipage}%
\hfill%
\begin{minipage}{0.85\textwidth}%\raggedleft
{\footnotesize \sf This work was supported by the European Union's Horizon 2020
research 
and innovation programme under the Marie Sk\l{}odowska-Curie grant 
agreement No.~764850 {\it ``\href{https://sagex.org}{SAGEX}''}.}
\end{minipage}

\newpage

%\noindent
\hspace{1.7cm}{\bf Affiliations:}\\

\address{$1$ Centre for Theoretical Physics, 
						Department  of Physics and Astronomy,\\
						Queen Mary University of London, 
						London E1 4NS,  United Kingdom}
						\vspace{3pt}
						
						\address{$2$
				Department of Mathematical Sciences, Durham University, Durham DH1 3LE, United Kingdom}
				\vspace{3pt}

\address{$3$ School of Mathematics and Hamilton Mathematics Institute, Trinity College Dublin,
Dublin 2, Ireland}
\vspace{3pt}
					
							\address{$4$ Institut f\"ur Physik und IRIS Adlershof, Humboldt-Universit\"at zu Berlin, \\
  Zum Gro{\ss}en Windkanal 2, D-12489 Berlin, Germany}
  \vspace{3pt}

			\address{$5$ Theoretical Physics Department, CERN, 1211 Geneva, Switzerland}
			
			\vspace{3pt}

			\address{$6$ Higgs Centre for Theoretical Physics, School of Physics and Astronomy,\\
The University of Edinburgh, Edinburgh EH9 3FD, United Kingdom}
\vspace{3pt}

\address{$7$ Mani L.~Bhaumik Institute for Theoretical Physics, Department of Physics and Astronomy, UCLA, Los Angeles, CA 90095, USA}

\vspace{3pt}

	\address{$8$ Niels Bohr International Academy, Niels Bohr Institute, University of Copenhagen, Blegdamsvej 17, DK-2100 Copenhagen, Denmark}
	
\vspace{3pt}

			\address{$9$
			Deutsches Elektronen–Synchrotron DESY,
			Platanenallee 6, 
			D-15738~Zeuthen,~Germany}
\vspace{3pt}

\address{$10$ Department of Physics and Astronomy, Northwestern University, Evanston, Illinois 60208, USA}

\vspace{3pt}

\address{$11$ Institut de Physique Th\'{e}orique, Universit\'{e} Paris Saclay, CEA, CNRS, F-91191 Gif-sur-Yvette Cedex, France}

\vspace{3pt}

\address{$12$ LAPTh, Universit\'{e} Savoie 
Mont Blanc, CNRS, B.P. 110, F-74941 Annecy-le-Vieux, France}
\vspace{3pt}

\address{$13$ Department of Physics and Astronomy, Uppsala University, Box 516, 75120 Uppsala, Sweden}
\vspace{3pt}

\address{$14$ Institute for Theoretical Physics, ETH Z\"{u}rich, 8093 Z\"{u}rich, Switzerland}
\vspace{3pt}

\address{$15$ Physik-Institut,  Universit\"{a}t Z\"{u}rich, Winterthurerstrasse 190, 8057 Z\"{u}rich, Switzerland}
\vspace{3pt}

\address{$16$ On leave from INFN, Laboratori Nazionali di Frascati, Italy}
\vspace{3pt}

				\address{$17$
				SLAC National Accelerator Laboratory, Stanford University, Stanford, CA 94309, USA}
				\vspace{3pt}

\address{$18$ Bethe Center for Theoretical Physics, Universit\'{a}t Bonn, D-53115, Germany}
				\vspace{3pt}
				
				\address{$19$ Department of Physics, Faculty of Science, Chulalongkorn University Thanon Phayathai, Pathumwan, Bangkok 10330, Thailand}
\vspace{3pt}

\address{$20$ Department of Applied Mathematics and Theoretical Physics, Wilberforce Road, Cambridge CB3 0WA, United Kingdom}
		\vspace{3pt}
	
	\address{$21$ Nordita, Stockholm University and KTH Royal Institute of Technology,  Hannes Alfv\'{e}ns  v\"{a}g 12, 10691 Stockholm, Sweden}
	\vspace{3pt}
	
	\address{$22$ Institut des Hautes \'{E}tudes Scientifiques, F-91440 Bures-sur-Yvette, France}
	\vspace{3pt}

		\address{$23$
		The Mathematical Institute,
University of Oxford, Oxford OX1 3LP, United Kingdom}
	\vspace{3pt}

 	\address{$24$ Deutsches Elektronen-Synchrotron DESY, Notkestra{\ss}e 85, D-22607 Hamburg, Germany}
 \vspace{3pt}
 
  \address{$25$ CPHT, CNRS, \'{E}cole Polytechnique, IP Paris, F-91128 Palaiseau, France}
  \vspace{3pt}
  
  \address{$26$  Perimeter Institute for Theoretical Physics,
31 Caroline Street North, Waterloo, Ontario, Canada N2L 2Y5}
 \vspace{3pt}
  
  \address{$27$ Institute for Gravitation and the Cosmos,
Pennsylvania State University, University Park, PA 16802,~USA}
  \vspace{3pt}
  
  \address{$28$ Johannes Kepler University Linz, Research Institute for Symbolic Computation (RISC), Altenberger Straße 69, A-4040 Linz, Austria}
  \vspace{3pt}

  \address{$29$ Center for Quantum Mathematics and Physics (QMAP),
Department of Physics, University of California, Davis, CA 95616, USA}
\vspace{3pt}

%\ead{a.brandhuber@qmul.ac.uk, jan.plefka@hu-berlin.de, %g.travaglini@qmul.ac.uk}
%\vspace{0.5cm}
%\begin{indented}
%\item[]December 2022
%a M
%\end{indented}

%\newpage 

\vspace{1cm}

\tableofcontents

\newpage

\section{Foreword}

Scattering experiments have been key  in shaping our understanding of the laws 
of Nature and testing the consistency 
of our physical theories, from Rutherford's discovery of the atomic nucleus to the discovery of 
the Higgs boson at the Large Hadron Collider (LHC). 
In quantum scattering experiments only the probabilities of the possible outcomes can be predicted. 
These probabilities are in turn given by the absolute value squared of 
the quantum transition amplitudes -- or {\em scattering amplitudes} -- between initial and final states.

Sparked by remarkable simplicity  emerging at the end of extensive calculations, 
a wave of breakthroughs established  modern methods for scattering amplitudes as a major new 
field in theoretical high-energy physics. The demands for precise theoretical predictions for present and future experiments, as well as theorists' thirst for ever-more-stringent tests of their theoretical ideas, fuelled the development 
of highly efficient  methods which not only compute amplitudes but also 
extract physical quantities from them.

Calculations made possible by these new methods revealed remarkable  connections between theories relevant to particle scattering at 
the LHC, and General Relativity. These connections reach beyond scattering, and 
give a new perspective on black holes and the physics of gravitational waves 
observed by the LIGO and Virgo collaborations.
They paint a picture in which quantum field theory is formulated without reference to quantum fields, and  concepts such as unitarity and causality, and space-time itself, may emerge from yet to be discovered more fundamental principles.
From the interpretation of the signal in current LHC experiments 
to the understanding of graviton  scattering relevant to the earliest times in the universe, previously prohibitive calculations revealed extraordinary simplicity and structure, and kick-started the next wave of advances.

{\it
What is the source of this simplicity?  
Can it be exploited to trivialise quantum predictions?  
How can it confront the empirical challenges facing fundamental physics?} 
{\it How do our classical notions of space, time, and causality emerge from the  underlying quantum description of Nature?}

These profound questions drive our field and  represented a core inspiration for the  consortium
{\bf \href{https://sagex.org}{SAGEX}}, or {\bf Scattering Amplitudes: from Geometry to Experiment}, 
an \href{https://cordis.europa.eu/project/id/764850}{Innovative Training Network of the European Commission.}
Combining strengths from   varied 
methodologies, SAGEX  has been  unified by pursuing scattering amplitudes as the essential
quantities underlying our understanding of particle physics.  
This collection of review articles, which grew out of the activities of the consortium, aims to give 
an overview of recent developments in the modern scattering amplitudes programme, 
thriving on the synergy between  mathematics and  theoretical physics, guided by explicit calculations
that provide new theoretical data and expose new structure.
It represents  the work of researchers who have discovered tools to compute amplitudes in gauge theory and gravity;
mathematicians who are experts in twistor theory, algebraic number theory, integrability and Hopf algebras; 
computer scientists working on symbolic computation; 
and phenomenologists making cutting edge predictions for LHC experiments.
It is our hope that the next generation of researchers, some of whom may benefit from this review, will embrace
our quest to uncover the unifying principles underlying scattering and, more generally, 
quantum field theory, and apply 
them to enrich our knowledge of Nature.

If you wish to refer to this review as a whole, we suggest that you cite this introductory article only, with the  format: \\
\begin{center}
G.~Travaglini, A.~Brandhuber, P.~Dorey, T.~McLoughlin, eds., 
\\
{\it ``The SAGEX Review on Scattering Amplitudes''}, \\
\href{https://iopscience.iop.org/article/10.1088/1751-8121/ac8380}{J.~Phys.~A~{\bf 55} no.~44,  (2022) 443001},  \href{https://arxiv.org/abs/2203.13011}{\tt arXiv:2203.13011  }
\end{center}

\section{Invitation to read}

The following  sections  give a flavour of what the various chapters describe.

\subsection[Modern fundamentals of amplitudes]{Modern fundamentals of amplitudes \cite{Brandhuber:2022qbk}}

Scattering amplitudes are much simpler than what one would expect based on  Feynman diagrams. In this introductory chapter we undertake  two tasks. The first is to introduce the framework which makes the simplicity of amplitudes manifest --  the spinor-helicity formalism, which we describe in detail. The  colour decomposition of amplitudes further  allows us to focus on  objects -- the colour-ordered partial amplitudes -- which only depend on kinematic data, which we express using spinor  variables.  
The second  is to 
describe efficient methods to compute amplitudes avoiding completely Feynman diagrams, and uncover new symmetries which may help to explain this simplicity. At tree level we discuss the Britto-Cachazo-Feng-Witten recursion relation and apply it  to derive  the infinite sequence of Maximally Helicity Violating gluon amplitudes  in Yang-Mills theory, 
\begin{align}
A_n^{\rm MHV}(1^+,\dots, i^-, \dots, j^-, \dots,  n^+)=ig^{n-2}\, \frac{\langle{ij\rangle}^4}{\langle 12\rangle\langle 23\rangle \cdots  \langle n1\rangle}\, .
\nonumber
\end{align}
At loop level, we introduce the modern unitarity method, and apply it to the computation of several one-loop amplitudes, with and without supersymmetry. We explain how  symmetries appear in the context of  amplitudes: these can be manifest, such as  Poincar\'{e} and (super)conformal, or mysteriously hidden, such as the  dual superconformal and Yangian symmetries of the $S$-matrix of  $\mathcal{N}{=}4$ supersymmetric Yang-Mills theory. We also describe in detail universal factorisation theorems for  tree-level amplitudes,  in particular  soft and collinear limits, and give a brief excursus on form factors -- quantities which are partially off shell -- which can also be efficiently computed using modern on-shell methods.

\subsection[An invitation to colour-kinematics duality and the double copy]{An invitation to colour-kinematics duality and the double copy \cite{Bern:2022wqg}}

In this chapter, we present a mini-review of the double copy and its
underlying duality between colour and kinematics, focusing on scattering
amplitudes. With an appropriate rearrangement, gauge-theory scattering
amplitudes can be converted into gravitational ones via a simple
replacement of colour factors with kinematic expressions.  This
approach simplifies higher-order gauge and gravity calculations and
poses tantalising questions related to a unified framework underlying
relativistic quantum theories. While originally discovered in the
context of gauge and gravitational theories, we now understand the
double-copy structure to relate a variety of effective field theories
such as chiral perturbation theory and the Born-Infeld model, as well
as higher-derivative corrections.  Nontrivial applications of the
double copy include classification of supergravity theories, high-loop
ultraviolet properties of gravity theories, and, more recently,
gravitational-wave physics relevant to ongoing and future
observations.

After a brief summary on colour-kinematics duality and the double copy,
this chapter presents two distinct applications.  The basics are first
given for the simplest cases of scattering amplitudes, especially
where all particles are in the adjoint representation.  Comments are
included on various generalisations. An updated version of the web of
interconnected double-copy-constructible theories is given as a first
application. The second is the recent application of the double copy
to gravitational-wave physics.

\subsection[Mathematical structures in Feynman integrals]{Mathematical structures in Feynman integrals \cite{Abreu:2022mfk}}

Feynman integrals are ubiquitous in any calculation in perturbative quantum field theory. 
They are famously hard to compute, often presenting a bottleneck in the evaluation of higher-order
corrections for quantities of physical interest. The rich mathematical structure of dimensionally-regulated Feynman integrals, 
however, has led to the development of powerful new techniques for their computation. 
In this review, we discuss some of the recent progress that has been made in understanding the mathematical structure
of Feynman integrals and in developing new approaches for their evaluation. 
After a brief introduction on the properties and different representations of Feynman integrals,
we discuss how linear relations between Feynman integrals arise and how they can be used to define
bases for classes of integrals, how differential equations can be used
to compute Feynman integrals but also to expose their mathematical structure, and how iterated integrals
are the natural class of functions that arises in the Laurent expansion of dimensionally-regulated Feynman integrals.
Finally, we briefly review intersection theory and the theory of twisted co/homology groups, focusing on how it provides
the natural mathematical interpretation for dimensional regularisation, putting many of the well-known properties of
Feynman integrals inside a single consistent mathematical framework. 
Throughout the review, we illustrate the different topics we discuss with simple examples.

\subsection[Multi-loop Feynman integrals]{Multi-loop Feynman integrals \cite{Blumlein:2022zkr}}

The analytic integration and simplification of multi-loop Feynman
integrals to special functions and constants play an important role in performing higher
order perturbative calculations in the Standard Model of elementary particles. In this chapter, 
the most recent and relevant computer algebra and special function
algorithms are presented that are currently used or that may play an important
role to perform such challenging precision calculations in the future. Among them,
we will focus on (1) guessing methods for the discovery of relations among special constants
and the calculation of linear recurrences and differential equations, (2) solving methods for
computing closed form solutions of (coupled systems of ordinary and partial) linear difference and differential 
equations,
and (3) simplification methods, like symbolic summation and integration for the explicit treatment of
multi-loop Feynman integrals and their related multi-sums. One of the limits of all these methods
is the observation that the complexity of the function spaces increase the more involved Feynman integrals, with respect to
the number of scales and loops, are tackled.
To overcome such difficulties, we present in addition the recently developed large moment method that features the
distinguished property that during intermediate calculation steps no functions arise and in the end only those 
functions
appear that are necessary to obtain optimal representations of the final physical result.
Furthermore, the arising function spaces and symbolic algorithms for their manipulation and
exploration are introduced that arise in the results of multi-loop Feynman calculations.
We conclude this chapter by presenting concrete analytic computations in the quantum field theories of the 
Standard Model at the massless and massive three-loop level in QCD and QED, including two mass scales in part.
Applications within effective field theories that utilise all the different solving and simplification tools 
introduced in this section, as notably the post-Newtonian expansion of classical (non-linear) Einstein gravity, 
are given to five-loop order.

\subsection[Analytic bootstraps for scattering amplitudes and beyond]{Analytic bootstraps for scattering amplitudes and beyond \cite{Papathanasiou:2022lan}}  

One of the main challenges in obtaining predictions for collider experiments from quantum field theory, is the evaluation of the Feynman integrals resulting from its traditional perturbative treatment. In this chapter, we review an alternative bootstrap method that bypasses this formidable task by constructing physical quantities from the knowledge of their expected analytic structure. Originally developed for six- and seven-particle amplitudes in the large-colour limit of the simplest interacting gauge theory, known as $\mathcal{N}{=}4$ super Yang-Mills (SYM), apart from their computation to unprecedented loop orders it has also revealed potentially universal properties of quantum field theory, as well as found application in various other different contexts.

After briefly introducing the $\mathcal{N}{=}4$ SYM amplitudes of interest, we move on to discuss the general class of polylogarithmic functions they belong to. We describe how mathematical objects known as cluster algebras provide strong clues for the singularities of these amplitudes at multiplicities $n{=}6,7$, enabling their bootstrap by rendering the aforementioned space of functions finite at each loop order. We also highlight the additional information cluster algebras are found to encode on how these singularities can appear consecutively, closely related to a generalisation of the Steinmann relations of axiomatic quantum field theory, which drastically reduces the size of the function space, and thus correspondingly simplifies its construction. We then explain the steps of this construction and of the unique identification of the amplitude within the function space with input from its expected behaviour in kinematic limits, and apply the general procedure to a concrete example: the determination of the two-loop correction to the first nontrivial six-particle amplitude.

Finally, we overview some of the current frontiers of the perturbative analytic bootstrap methodology. These include a proposed resolution of the longstanding difficulties in identifying the appropriate cluster algebra generalisations dictating the singularities of amplitudes at higher multiplicity $n{>}7$, as well as the application to the computation of other quantities such as form factors and Feynman integrals.

\subsection[Ambitwistor strings and amplitudes from the worldsheet]{Ambitwistor strings and amplitudes from the worldsheet \cite{Geyer:2022cey}}

From its beginning, string theory has highlighted otherwise hidden worldsheet structures underlying scattering amplitudes although for better or worse, it is a theory that goes far beyond conventional field theory.  Much more recently, starting with Witten's twistor string, chiral string theories have emerged that  yield    \emph{field theory} amplitudes alone without the extra towers of massive modes of conventional strings. These string theories do so very directly and yield compact formulae for amplitudes at tree level and more recently for loop integrands.
These worldsheet models have as target the space of complexified null geodesics,  known as ambitwistor space, and hence are referred to as ambitwistor strings.  They naturally  connect the amplitudes to a worldsheet, the Riemann sphere at tree-level, localising on residues supported on solutions to the \emph{scattering equations}. One family of models gives the theory underpinning to the worldsheet formulae of Cachazo, He and Yuan (CHY) whereas others lead to twistorial formulae such as those of Roiban, Spradlin and Volovich.
In this chapter, we discuss two incarnations of the ambitwistor string:  a ``vector representation'' starting in space-time structurally resembling the bosonic and RNS  superstring, and a four-dimensional twistorial version closely related to, but distinct from Witten's original model.
The RNS models exist for several theories, with ``heterotic'' and type II models describing super Yang-Mills and ten-dimensional supergravity,  respectively. They  manifest the double-copy relations between gauge theory and gravity, phrased here directly on the worldsheet, and correlators of vertex operators give the remarkable CHY formulae for the respective theories. 
%At tree-level, this expresses the $S$-matrix as an integral over a marked Riemann sphere --  the worldsheet -- with  punctures associated to the scattered particles. Importantly however, these integrals are fully localised on a set of constraints known as the  \emph{scattering equations} that play a pivotal role in relating the moduli space of the worldsheet to the kinematic configuration space of massless particles. 
In the remainder of the chapter, we utilise the advantage provided by the underlying model to showcase diverse applications, ranging from loop amplitudes to scattering on curved backgrounds, and interesting connections to celestial holography.

\subsection[Positive geometry of scattering amplitudes]{Positive geometry of scattering amplitudes \cite{Herrmann:2022nkh}}

% \blue Half a page per summary, max one equation. \black

In this chapter, we summarise some of the recent developments in rephrasing perturbative scattering amplitudes in an entirely novel geometric way, where traditional notions such as locality and unitarity are secondary concepts derived from fundamental mathematical properties of positive geometries. We discuss these geometric ideas on two concrete examples: First for tree-level scattering amplitudes in bi-adjoint scalar $\phi^3$ theory and their connection to the associahedron polytope with its boundaries of all co-dimensions. The physical factorisation properties of scattering amplitudes are entirely encoded in the factorisation properties of the associahedron on those boundaries. Our second example is the amplituhedron construction for tree-level amplitudes and the all-loop integrand in planar maximally supersymmetric Yang-Mills theory. This is a more intricate geometry closely related to the positive Grassmannian $G_+(k,n)$ and the combinatorics of on-shell diagrams. While the associahedron geometry is naturally defined in the space of Mandelstam variables, the amplituhedron lives in momentum twistor space. We close our chapter with a brief account of other areas of theoretical physics where positive geometries have recently played a role, such as  cosmological correlation functions, the conformal bootstrap, bounds on low-energy EFT Wilson coefficients via the EFT-Hedron construction, as well as the definition of the amplituhedron in spinor helicity space, and some attempts to extend the geometric construction beyond the planar limit.

\subsection[Half BPS correlators]{Half BPS correlators \cite{Heslop:2022qgf}}

$\cN{=}4$ supersymmetric Yang-Mills (SYM) is the   most symmetric of all  four-dimensional quantum field theories. The simplest local operators in this theory are the half BPS operators. Despite this simplicity, the family of correlators of half BPS operators provides a remarkably  rich breeding ground of ideas  as well as being some of the most accurately known quantities in any four-dimensional quantum field theory. 
 They  impact  on many key areas in current theoretical physics such as  scattering amplitudes, integrability, positive geometry/amplituhedron and the  conformal bootstrap. 

The  relation of half BPS correlators to scattering amplitudes is of particular interest and importance and has provided the thrust for most of the progress. This appears in two completely different ways, to amplitudes in two different theories. 
They  first became significant objects of interest immediately after the discovery in 1997 of the AdS/CFT correspondence  which  tells us that we can interpret them as  IIB supergraviton amplitudes in string theory on an AdS${}_5{\times} S^5$ background. 
 Indeed the most accurate quantum gravity amplitudes in curved space has recently been obtained by bootstrapping a half BPS correlator at strong coupling  and this provides a crucial arena for exploring quantum gravity.
But then over a decade after AdS/CFT in 2010 another relation to amplitudes was discovered, the correlator/amplitude duality.
 The correlator/amplitude duality states that the correlators become $\cN{=}4$ SYM amplitudes on taking a certain polygonal lightlike limit. This duality  gives insight in both directions and indeed the most accurately known amplitude integrand  has been obtained via the correlator using this duality.

In this review we will attempt to describe as much as possible of what is currently known of the family of half BPS correlators in $\cN{=}4$ SYM.

%\subsection[Integrability for %Amplitudes -- The Pentagon %OPE]{Integrability for Amplitudes -- %The Pentagon OPE \cite{chapter9}}
%
%\blue Half a page per summary, max one %equation. \black

\subsection[Integrability of amplitudes in 
fishnet theories]{Integrability of amplitudes in 
fishnet theories \cite{Chicherin:2022nqq}}

%\blue Half a page per summary, max one equation. 
The four-dimensional maximally supersymmetric Yang-Mills theory has been serving for several decades as a theoretical laboratory for developing novel techniques for computing scattering amplitudes in realistic gauge theories. 
As a result of partial cancellations among contributions from particles of different spins, the scattering amplitudes in this theory have a remarkably simple form. This property is a manifestation of integrability of the theory and it 
is completely obscured in the Lagrangian formulation. 

The conformal fishnet theories arise as certain integrable deformations of the maximally supersymmetric Yang-Mills theory. One of the motivations for studying these theories 
is that they allow us to elucidate the origin of integrability. The simplest bi-scalar fishnet theory describes two complex scalar particles with a peculiar quartic interaction endowed with chirality. The latter property ensures that 
the planar scattering amplitudes receive contributions only from fishnet-like Feynman diagrams and renders the theory integrable in the planar limit for an arbitrary value of the coupling. In this chapter, we review the all-loop calculation 
of the four-particle scattering amplitude in the conformal fishnet theory and discuss the Yangian symmetry of the multi-particle scattering amplitudes.

\black

\subsection[Modular covariance of type IIB string amplitudes
and their $\mathcal{N} {=}4$ supersymmetric Yang-Mills duals]{Modular covariance of type IIB string amplitudes
and their $\mathcal{N} {=}4$ supersymmetric Yang-Mills duals \cite{Dorigoni:2022iem}}

This chapter focuses on the interplay of maximal supersymmetry and modular covariance in determining scattering amplitudes in type IIB superstring theory and the holographic dual of these amplitudes, which are correlation functions in $\mathcal{N} {=}\, 4$ supersymmetric Yang-Mills theory (SYM).  We demonstrate how the combination of these symmetries is used to determine  certain exact properties of superstring scattering amplitudes  and SYM correlators. 

 In the first part we discuss  how  supersymmetry and modular covariance imply strong constraints that determine coefficients of BPS terms in the low energy expansion of  type IIB superstring amplitudes. We elucidate, in particular, exact properties of $n$-point amplitudes with $n\ge 4$ that violate $U(1)$  R-symmetry maximally.
 
  In the second part we demonstrate that certain integrated correlators  of gauge invariant operators in the stress tensor supermultiplet  in $\mathcal{N} {=}\, 4$ SYM  can be computed exactly using supersymmetric localisation techniques.  This analysis applies to SYM with any classical gauge group, $G_N=SU(N)$, $SO(2N)$, $SO(2N+1)$ or $USp(2N)$.  We exhibit a variety of intriguing properties of perturbation theory at any fixed value of $N$ as well as  properties of the large-$N$ expansion, which make contact with the low-energy expansion of type IIB superstring amplitudes in AdS${}_5{\times} S^5$ and AdS${}_5{\times} S^5/{\mathbb{Z}}_2$.   
 
 In the final part, we focus on modular graph functions. These are modular functions that are closely associated with coefficients in the low-energy expansion of  superstring perturbation theory. Their properties reflect the interplay of supersymmetry and the modular properties associated with diffeomorphisms of string world-sheets.  For the most part we concentrate on the $\alpha' {\to} 0$ expansion of genus-one world-sheets amplitudes, where certain modular graph functions are related to interesting generalisations of
non-holomorphic Eisenstein series.  

These topics involve modern developments at the interface of quantum field theory,  quantum gravity and string theory, and have surprising interconnections with modern areas of algebraic geometry and number theory.

\subsection[Soft theorems and celestial amplitudes]{Soft theorems and celestial amplitudes \cite{McLoughlin:2022ljp}}

In this chapter, we first review tree-level soft theorems for photons, gluons and gravitons as well as their loop-corrections and multi-soft generalisations. We then briefly discuss the connection to asymptotic, or large, gauge symmetries and the reformulation of soft theorems as Ward identities for the corresponding conserved currents. Recasting the sub-leading soft graviton theorem as the Ward identity for a two-dimensional stress-tensor naturally leads to considering the $S$-matrix in the basis of boost eigenstates. The $S$-matrix elements then have a natural interpretation as correlation functions in a holographically dual \textit{celestial} conformal field theory (CCFT). We review the definition of celestial operators in terms of conformal primary wavefunctions and the construction of celestial amplitudes by means of integral transforms. We provide an overview of the structure of these amplitudes, including their analytical properties, OPE-like behaviour in collinear limits and realisation of the double-copy construction. 

Asymptotic symmetry charges are related to  currents in the boundary CCFT which in turn control the behaviour of celestial amplitudes in conformally soft limits. We study the Ward identities these conformally soft currents imply and the constraints they impose on the form of celestial OPEs. Examining the conformal multiplets of CCFT reveals the existence of infinitely many currents beyond those with a clear asymptotic symmetry interpretation. The towers of conformally soft currents can be shown to obey infinite dimensional algebras which in the case of gravity is the wedge algebra of $w_{1+\infty}$. Finally, we discuss the relation of exponentiated IR divergences in scattering amplitudes to correlation functions of vertex operators for Goldstone bosons in CCFT and the conformal analogue of Faddeev-Kulish dressed states.

\subsection[Amplitudes and collider physics]{Amplitudes and collider physics \cite{White:2022wbr}}

In this chapter, we examine one of the main applications of scattering
amplitudes, namely the calculation of predictions for collider
experiments such as the ongoing Large Hadron Collider (LHC). We first
review what colliders are and why they are useful, before describing
some of the main quantities that are measured in these experiments
e.g. (differential) cross-sections. We explain why the simple notion
of comparing ``theory'' to ``data'' is in practice a lot more
complicated than it sounds, and we also take some topics that are of
particular interest to more formal theorists, and show that these have
very practical applications in improving the precision of our
theoretical predictions.

The topics covered include how to describe collisions with incoming
protons rather than quarks/gluons, and what the state-of-the-art of
predictions for interesting LHC processes is. We then examine how, for
some observables, we must sum up contributions to all orders in
perturbation theory. This process is known as {\it resummation}, and
turns out to be related to the study of {\it (next-to)-soft} radiation
i.e.~the emission of quarks and/or gluons whose energy and momentum
is low. We outline the detailed procedures that are needed to take a
theoretical calculation, and get it closer to something that looks
like what happens in a real particle accelerator. We also discuss the
various ways in which theory can be meaningfully compared to data,
some more complicated than~others!

\subsection[Post-Minkowskian expansion from
scattering amplitudes]{Post-Minkowskian expansion from
scattering amplitudes \cite{Bjerrum-Bohr:2022blt}}

The Post-Minkowskian expansion in general relativity is based on expanding observable quantities in Newton's constant $G_N$ only. Measuring physical quantities at infinity where space-time is Minkowskian, this corresponds to an expansion in gravity that is special relativistic (valid to all orders in velocities) and without the imposition of the counting based upon the virial theorem for bound orbits. Relativistic quantum field theory is the ideal set-up for this situation even though it also includes all the quantum-mechanical effects that must be discarded when seeking predictions for classical physics. This set-up is most naturally implemented in the case of gravitational scattering, from which the effective gravitational interactions order by order in $G_N $ can be inferred.

A number of different methods are now available for this extraction of classical gravitational physics from  amplitudes. With each new order in the expansion, new issues arise and it is therefore important to search for the most efficient method. Here we shall focus on the case of non-spinning compact objects scattering by gravity only, the conceptually simplest case. A central issue is how to proceed from $S$-matrix elements to observable quantities such as classical scattering angles. Surprisingly, we find that quantum mechanical unitarity plays a crucial role in this process: from the evaluation of the actual scattering amplitudes to the understanding of which subtractions must be introduced in order to render the final result classical. This is especially clear in a recent formulation based on an exponential representation of the $S$-matrix.

At the third Post-Minkowskian order in the expansion, the classical part of the scattering amplitude already includes effects that can be understood as radiation reaction pieces, a phenomenon first identified by consistency conditions on the amplitude. We show how these radiation reaction terms arise straightforwardly from the amplitude when all classical terms are extracted from the integrations. At the same time, this shows full agreement between the eikonal formalism and the formulation based on the exponential representation of the amplitude. 

\subsection[Classical gravity from scattering amplitudes]{Classical  gravity from scattering amplitudes \cite{Kosower:2022yvp}}

Scattering amplitudes capture basic on-shell physical information
in a beautifully simple form.
Although the amplitudes have their origin in quantum field theory,
they have wide-ranging applications even in classical physics.
In this review article, we discuss how to compute certain observables
from scattering amplitudes, focusing on scattering events and on observables that have a
well-defined classical limit.
These observables include the impulse on a particle (that is, its
total change in momentum);
the momentum radiated via the electromagnetic or the
gravitational field; and gravitational waveforms.
As these observables can be expressed in terms of scattering
amplitudes, it follows directly that the double copy can be used
to construct classical gravitational observables from corresponding
Yang-Mills amplitudes.

The double copy is, of course, at its simplest for
three-point amplitudes.
We review how three-point amplitudes, and the double copy, compute the
linearised curvature of certain stationary solutions in general
relativity after analytically continuing to metric
signature $(+,+,-,-)$.
Building on this construction, we demonstrate explicitly that
these linearised gravitational objects are double copies of
electromagnetic counterparts.
Moving beyond linearised theory, it is remarkable that important
classes of exact gravity solutions admit a straightforward
interpretation as a double copy of solutions in electromagnetism. We
discuss the Weyl double copy, emphasising its firm grounding in scattering amplitudes,
and connect it to Kerr-Schild metrics.

\subsection[The Multi-Regge limit]{The Multi-Regge limit \cite{DelDuca:2022skz}}

At high-energy hadron colliders, jets of hadrons are produced copiously,
emerging from underlying collisions of quarks and gluons, which can have
high multiplicity due to the nonabelian nature of quantum chromodynamics (QCD).
Multi-jet production extends over large ranges in rapidity
(roughly the logarithm of the polar angle), and is further enhanced
logarithmically in such Regge limits and multi-Regge kinematics (MRK).

This chapter will explore our theoretical understanding of these limits
in both QCD and its maximally supersymmetric cousin, planar ${\cal N}{=}4$
super Yang-Mills (SYM) theory.  Perturbative scattering amplitudes in these theories
factorise in such limits.  Large logarithms are associated with
Reggeized gluons and their associated trajectories, which are computable
via the BFKL equation.  The most forward
and backward part of the process are described by impact factors,
while central-emission vertices characterise the production of gluons
at intermediate rapidities.  Factorisation and BFKL evolution
is most transparent, not in momentum space, but after performing a
combined Fourier and Mellin transformation.

While the factorised structure is reasonably well understood
in QCD, one can push much further in planar ${\cal N}{=}4$
SYM theory, an integrable theory.  Integrability
has been exploited
to propose formulae for the BFKL eigenvalue, the impact factor,
and more recently the central-emission vertex, formulas which should be
valid to all orders in perturbation theory.  These proposals have
been checked via independent computations of six-gluon amplitudes
through seven loops, and (for the central-emission vertices) seven-gluon
amplitudes through four loops.

At each order in perturbation theory, the analytic functions encountered
are single-valued, or real-analytic,
multiple polylogarithms of $n{-}5$ complex variables
associated with the complexified transverse-momentum plane.
Although these functions are fairly intricate, they are still considerably
simpler than those required to describe scattering in general kinematics,
making the multi-Regge limit a unique window into the deeper mathematical
structure of scattering amplitudes.

%\newpage
\vspace{1cm}

\section*{Acknowledgments}

\addcontentsline{toc}{section}{\protect\numberline{}Acknowledgments}
We would like to express our warmest thanks to Jenna Lane and Mary Thomas, who have given invaluable help over the years to {\it \href{https://sagex.org}{SAGEX}} and in particular to the coordination of this review.  
This work  was supported  by the European Union's Horizon 2020 research and innovation programme under the Marie Sk\l{}odowska-Curie grant agreement No.~764850 {\it ``\href{https://sagex.org}{SAGEX}''}.   \\

\bibliographystyle{utphys}
\newcommand{\eprint}[2][]{\href{https://arxiv.org/abs/#2}{\tt{#2}}}
\addcontentsline{toc}{section}{\protect\numberline{}References}
\section*{References}

\bibliography{review-biblio}

\end{document}